\documentclass[twocolumn,tighten]{aastex63}
\usepackage{natbib}
\usepackage{epsfig}
\usepackage{amsmath}
\usepackage{xcolor}
\usepackage{hyperref}
\usepackage{booktabs}
\usepackage{xtab}
\usepackage{afterpage}
\usepackage[encapsulated]{CJK}
\usepackage{ucs}
\usepackage[utf8x]{inputenc}
\usepackage{url}
\usepackage{lineno}
\usepackage[caption=false]{subfig}
\usepackage{array}

\usepackage{xspace}
\usepackage{comment}
\usepackage{threeparttable}
\newcommand{\name}{SN~2021aefx\xspace}
\newcommand{\mat}[1]{\textcolor{purple}{MAT: #1}} 

\newcommand{\um}{\hbox{\ensuremath{\mu\rm m}}\xspace}

\newcommand{\jwst}{\emph{JWST}\xspace}
\newcommand{\sneia}{SNe Ia\xspace}
\newcommand{\snia}{SN Ia\xspace}
\newcommand{\tmax}{\hbox{\ensuremath{t_{\rm{max}}}}\xspace}

\definecolor{maroon}{rgb}{0.760,0.118,0.337}

\begin{document}

\newcommand{\OSU}{\affil{Department of Astronomy, The Ohio State University, 140 West 18th Avenue, Columbus, Ohio 43210, USA}}

\newcommand{\CCAPP}{\affil{Center for Cosmology and Astroparticle Physics, 191 West Woodruff Avenue, Columbus, OH 43210, USA}}

\newcommand{\NRAO}{\affil{National Radio Astronomy Observatory, 520 Edgemont Road, Charlottesville, VA 22903-2475, USA}}

\newcommand{\UAlberta}{\affil{4-183 CCIS, University of Alberta, Edmonton, Alberta, T6G 2E1, Canada}}

\newcommand{\Rutgers}{\affiliation{Department of Physics and Astronomy, Rutgers, the State University of New Jersey,\\136 Frelinghuysen Road, Piscataway, NJ 08854, USA}}

\newcommand{\STScI}{\affiliation{Space Telescope Science Institute, 3700 San Martin Drive, Baltimore, MD 21218, USA}}

\newcommand{\Alberta}{\affil{Department of Physics, University of Alberta, Edmonton, AB T6G 2E1, Canada}}

\newcommand{\HD}{\affil{Astronomisches Rechen-Institut, Zentrum f\"{u}r Astronomie der Universit\"{a}t Heidelberg, M\"{o}nchhofstra\ss e 12-14, D-69120 Heidelberg, Germany}}

\newcommand{\ITA}{\affiliation{Universit\"{a}t Heidelberg, Zentrum f\"{u}r Astronomie, Institut f\"{u}r Theoretische Astrophysik, Albert-Ueberle-Str 2, D-69120 Heidelberg, Germany}}

\newcommand{\IWR}{\affiliation{Universit\"{a}t Heidelberg, Interdisziplin\"{a}res Zentrum f\"{u}r Wissenschaftliches Rechnen, Im Neuenheimer Feld 205, D-69120 Heidelberg, Germany}}

\newcommand{\ANU}{\affiliation{Research School of Astronomy and Astrophysics, Australian National University, Canberra, ACT 2611, Australia}}

\newcommand{\ASTROThreeD}{\affiliation{ARC Centre of Excellence for All Sky Astrophysics in 3 Dimensions (ASTRO 3D), Australia}}

\newcommand{\MPIA}{\affiliation{Max Planck Institute for Astronomy, K\"onigstuhl 17, 69117 Heidelberg, Germany}}

\newcommand{\MISSING}{\affiliation{\textcolor{purple}{Need affiliation}}}

\shorttitle{SN~2021aefx with JWST}
\shortauthors{Mayker~Chen et al.}


\title{Serendipitous Nebular-phase \emph{JWST} Imaging of SN~Ia~2021aefx: \\ Testing the Confinement of $^{56}$Co Decay Energy}

\begin{abstract}
We present new $0.3-21~\um$ photometry of \name in the spiral galaxy NGC~1566 at $+357$ days after $B$-band maximum, including the first detection of any SN Ia at $>15~\micron$. These observations follow earlier \jwst observations of \name at $+255$ days after the time of maximum brightness, allowing us to probe the temporal evolution of the emission properties. We measure the fraction of flux emerging at different wavelengths and its temporal evolution. Additionally, the integrated $0.3-14~\um$ decay rate of $\Delta m_{0.3-14} = 1.35 \pm 0.05 $~mag/100~days is higher than the decline rate from the radioactive decay of $^{56}$Co of $\sim 1.2$~mag/100~days. The most plausible explanation for this discrepancy is that flux is shifting to $>14~\um$, and future \jwst observations of \sneia will be able to directly test this hypothesis. However, models predicting non-radiative energy loss cannot be excluded with the present data. 
\end{abstract}

\keywords{interstellar medium: star-formation - supernova: locations - galaxies: nearby}

\correspondingauthor{Ness Mayker~Chen}
\email{maykerchen.1@osu.edu}

\author[0000-0002-5993-6685]{Ness Mayker~Chen}
\OSU \CCAPP

\author[0000-0002-2471-8442]{Michael A. Tucker}
\altaffiliation{CCAPP Fellow}
\CCAPP
\OSU 

\author[0000-0001-8040-4088]{Nils~Hoyer}
\affiliation{Donostia International Physics Center, Paseo Manuel de Lardizabal 4, E-20118 Donostia-San Sebasti{\'{a}}n, Spain}

\author{Saurabh W. Jha}
\Rutgers

\author{Lindsey Kwok}
\Rutgers

\author[0000-0002-2545-1700]{Adam~K.~Leroy}
\OSU\CCAPP

\author[0000-0002-5204-2259]{Erik Rosolowsky}
\Alberta

\author{Chris Ashall}
\affiliation{Department of Physics, Virginia Tech, Blacksburg, VA 24061, USA}

\author[0000-0002-5259-2314]{Gagandeep Anand}
\STScI

\author[0000-0003-0166-9745]{Frank Bigiel}
\affiliation{Argelander-Institut f\"ur Astronomie, Universit\"at Bonn, Auf dem H\"ugel 71, 53121 Bonn, Germany}

\author{Médéric Boquien}
\affiliation{Centro de Astronomía (CITEVA), Universidad de Antofagasta, Avenida Angamos 601, Antofagasta, Chile}

\author{Chris Burns}
\affil{Observatories of the Carnegie Institution for Science, 813 Santa Barbara Street, Pasadena, CA 91101, USA}

\author[0000-0002-5782-9093]{Daniel Dale}
\affiliation{Department of Physics and Astronomy, University of Wyoming, Laramie, WY 82071, USA}

\author[0000-0002-7566-6080]{James~M.~DerKacy}
\affiliation{Department of Physics, Virginia Tech, Blacksburg, VA 24061, USA}

\author[0000-0002-4755-118X]{Oleg V. Egorov}
\HD

\author[0000-0002-1296-6887]{L. Galbany}
\affiliation{Institute of Space Sciences (ICE, CSIC), Campus UAB, Carrer de Can Magrans, s/n, E-08193 Barcelona, Spain}
\affiliation{Institut d’Estudis Espacials de Catalunya (IEEC), E-08034 Barcelona, Spain}

\author[0000-0002-3247-5321]{Kathryn~Grasha}
\altaffiliation{ARC DECRA Fellow}
\ANU
\ASTROThreeD

\author[0000-0002-8806-6308]{Hamid Hassani}
\Alberta

\author{Peter Hoeflich}
\affil{Department of Physics, Florida State University, 77 Chieftan Way, Tallahassee, FL 32306, USA}

\author[0000-0003-1039-2928]{Eric Hsiao}
\affil{Department of Physics, Florida State University, 77 Chieftan Way, Tallahassee, FL 32306, USA}

\author[0000-0002-0560-3172]{Ralf S.\ Klessen}
\ITA
\IWR

\author[0000-0002-1790-3148]{Laura A. Lopez}
\OSU \CCAPP

\author[0000-0002-3900-1452]{Jing Lu}
\affil{Department of Physics, Florida State University, 77 Chieftan Way, Tallahassee, FL 32306, USA}

\author[0000-0003-2535-3091]{Nidia Morrell}
\affil{Las Campanas Observatory, Carnegie Observatories, Casilla 601, La Serena, Chile}

\author{Mariana Orellana}
\affiliation{Universidad Nacional de Río Negro. Sede Andina, Mitre 630 (8400), Bariloche, Argentina}
\affiliation{Consejo Nacional de Investigaciones Científicas y Tećnicas (CONICET), Argentina}

\author[0000-0001-5965-3530]{Francesca~Pinna}
\MPIA

\author[0000-0002-4781-7291]{Sumit K. Sarbadhicary}
\OSU \CCAPP

\author[0000-0002-3933-7677]{Eva Schinnerer}
\MPIA

\author[0000-0002-9301-5302]{Melissa Shahbandeh}
\affiliation{Department of Physics and Astronomy, Johns Hopkins University, Baltimore, MD 21218, USA}
\affiliation{Space Telescope Science Institute, 3700 San Martin Drive, Baltimore, MD 21218, USA}

\author[0000-0002-5571-1833]{Maximilian Stritzinger}
\affiliation{Department of Physics and Astronomy, Aarhus University, Ny Munkegade 120, DK-8000 Aarhus C, Denmark}

\author[0000-0002-8528-7340]{David A. Thilker}
\affiliation{Department of Physics and Astronomy, Johns Hopkins University, Baltimore, MD 21218, USA}

\author[0000-0002-0012-2142]{Thomas G. Williams}
\affiliation{Sub-department of Astrophysics, Department of Physics, University of Oxford, Keble Road, Oxford OX1 3RH, UK}



\section{Introduction}
\label{sec:Introduction}

Type Ia supernovae (SNe Ia) are the thermonuclear explosions of carbon-oxygen white dwarf stars \citep{Hoyle1960}. Their importance is two-fold: their use as standard candles is essential to the field of cosmology \citep[e.g., ][]{Schmidt1998}, and they produce the majority of iron-group elements in the Universe \citep[e.g., ][]{Iwamoto1999}. Despite their overarching importance, many open questions remain regarding their progenitor systems and explosion mechanisms (see \citealp{Maoz2014, Jha2019} for recent reviews).

\sneia are powered by the radioactive decay of unstable isotopes synthesized during the explosion \citep{Pankey1962}. The decay of
\begin{equation}
    ^{56}\mathrm{Ni}  \;\stackrel{t_{1/2} = \; 6.08d}{\hbox to 60pt{\rightarrowfill}} \; ^{56}\mathrm{Co} \; 
\stackrel{t_{1/2} = \; 77.2d}{\hbox to 60pt{\rightarrowfill}} \; ^{56}\mathrm{Fe}
\end{equation}
dominates the energy deposition into the ejecta for the first $\sim3$ years after explosion \citep[e.g., ][]{Seitenzahl2009}. $\gamma$-rays emitted during the first half of this decay chain, $^{56}\rm{Ni}\rightarrow ^{56}\rm{Co}$, provide the majority of the heating at $\lesssim 200$~days after explosion \citep[e.g., ][]{Arnett1982, Childress2015}. The second decay, $^{56}\rm{Co}\rightarrow{}^{56}\rm{Fe}$, produces X-ray and $\gamma$-ray photons in addition to high-energy ($\sim 1$~MeV) positrons ($e^+$) that dominate the energy deposition $\approx 200-1200$~days after explosion \citep[e.g., ][]{Kushnir2020, Tucker2022b}. 

For many years it was unclear if some of the ${^{56}\rm{Co}\rightarrow{}^{56}\rm{Fe}}$ decay energy escaped from the ejecta into the surrounding environment. Early studies of late-time \snia light curves found that the nebular-phase \citep[$\gtrsim 150$~days after maximum light, e.g.][]{Bowers1997, Branch2008} optical ($0.3-1~\um$) brightness declined faster than expected for the radioactive decay of $^{56}\rm{Co}$, suggesting a fraction ($\sim 1-10\%$) of the decay energy was not confined to the ejecta \citep[e.g., ][]{Milne2001}. This changed with the inclusion of $JHK$ ($1-2.5~\um$) observations which reconciled the (pseudo-)bolometric decline rate with expectations for pure $^{56}$Co decay \citep[e.g., ][]{Stanishev2007, Stritzinger2007, Leloudas2009}. 

A persistent uncertainty in these studies is the unknown fraction of flux emerging at $\gtrsim 2.5~\um$. \citet{Gerardy2007} published single-epoch nebular-phase \emph{Spitzer} spectra of \sneia 2003hv and 2005df extending to $15~\um$. \cite{Johansson2017} analyzed multi-epoch \emph{Spitzer} light curves of \sneia extending into the nebular phase, but were limited to the CH1 (3.6~\micron) and CH2 (4.5~\micron) bandpasses. \citet{Kwok2022} and \citet{derkacy2023} recently published spectra of \name extending to $14~\um$ with \jwst at +255 and +323~days after maximum light, respectively. SN~2014J has multiple mid-IR spectra \citep{Telesco2015} with the caveat that the spectra were obtained from the ground with the Gran Telescopio Canarias, where the high sky brightness and low atmospheric transmission in the mid-IR complicates precise flux calibration. To date, no \snia has been detected at $>15~\um$. Three \sneia (2011by, 2011fe, and 2012cg) were observed within 45 days of maximum light with \emph{Herschel} at 70~\um and 160~\um, but only upper limits were obtained \citep{Johansson2013b}. 

In this Letter, we report new optical and IR broad-band imaging of \name at +357~days after $B$-band maximum light. The new \jwst data were obtained as part of the Physics at High Angular Resolution of Nearby GalaxieS (PHANGS)-JWST Cycle 1 Treasury Program \citep{Lee2022}, which observed the host of \name, NGC~1566, in eight filters spanning $2-21~\um$ (\emph{F200W} to \emph{F2100W}). While the core science goals of PHANGS-JWST center on understanding stellar formation and feedback in nearby, star-forming galaxies, the serendipitous 2-21~$\micron$ observations of \name provide new insight into the cooling properties of \snia ejecta at nebular phases, including the first detection of a \snia at $>$~15~$\micron$.

\name was discovered on November 11, 2021 \citep[MJD 59529,][]{Valenti2021, Bostroem2021} by the Distance Less Than 40~Mpc (DLT40) transient survey \citep{Tartaglia2018}. Its host galaxy, NGC~1566, is a massive ($\log_{10} M_\star \left[ {\rm M}_\odot \right] \approx 10.8$), relatively face-on star-forming (SFR $\approx 4.5~{\rm M}_\odot~{\rm yr}^{-1}$) spiral galaxy \citep[][]{Leroy2021} 
at a distance $d=17.69 \pm 2.02$ Mpc (distance modulus $\mu = 31.24 \pm 0.27$ mag) \citep{Kourkchi2017,Anand2021} and redshift $z = 0.00502 \pm 0.00001$ \citep{Allison2014}. Despite excess $u$-band emission in the early light curve \citep{Ashall2022}, \name\ evolved into a normal SN Ia with $\Delta m_{15}(B) = 0.90 \pm 0.02$~mag and reached $M_B = -19.63 \pm 0.02$~mag on $t_{\rm max}$ = MJD 59546.5 \citep{Hosseinzadeh2022}. 

The brightness and close proximity of \name make it an excellent target for nebular-phase \jwst observations. \citet{Kwok2022} and \citet{derkacy2023} provided the first demonstration of the impressive spectroscopic capabilities of \jwst for studying nebular-phase \sneia. Their spectra of \name, obtained +255~days and +323~days after \tmax , respectively, represent the highest-quality look at the emission properties $>2.5~\um$ of SNe Ia to date. Their spectra show prominent emission features from the iron-group elements (Ni, Co, Fe), as well as a wide, flat-topped [\ion{Ar}{3}] profile that indicates a spherical shell of emission.

Our new $2-21~\um$ \jwst photometry at +357~days are used in conjunction with the +255~day $0.3-14~\um$ spectrum from \citet{Kwok2022} to 1) provide the time-dependent fraction of flux emerging at optical, near-IR, and mid-IR wavelengths; and 2) determine if the decline rate of the $0.3-14~\um$ flux is consistent with expectations for the radioactive decay of $^{56}\rm{Co}\rightarrow^{56}\rm{Fe}$. The imaging data and reduction are described in \S\ref{sec:Data}. \S\ref{sec:Results} outlines our analysis procedure, including constructing the 2$-$21$~\micron$ SED (Figure \ref{fig:SED}) and calculating filter-specific and bolometric decay rates (Figure \ref{fig:nebdec} and Table~\ref{tab:fluxes}). Finally, in \S\ref{sec:Discussion} we discuss and summarize our results.


\section{Observations}
\label{sec:Data}


\begin{figure*}
    \begin{center}   \includegraphics[width=0.8\linewidth, angle=270]{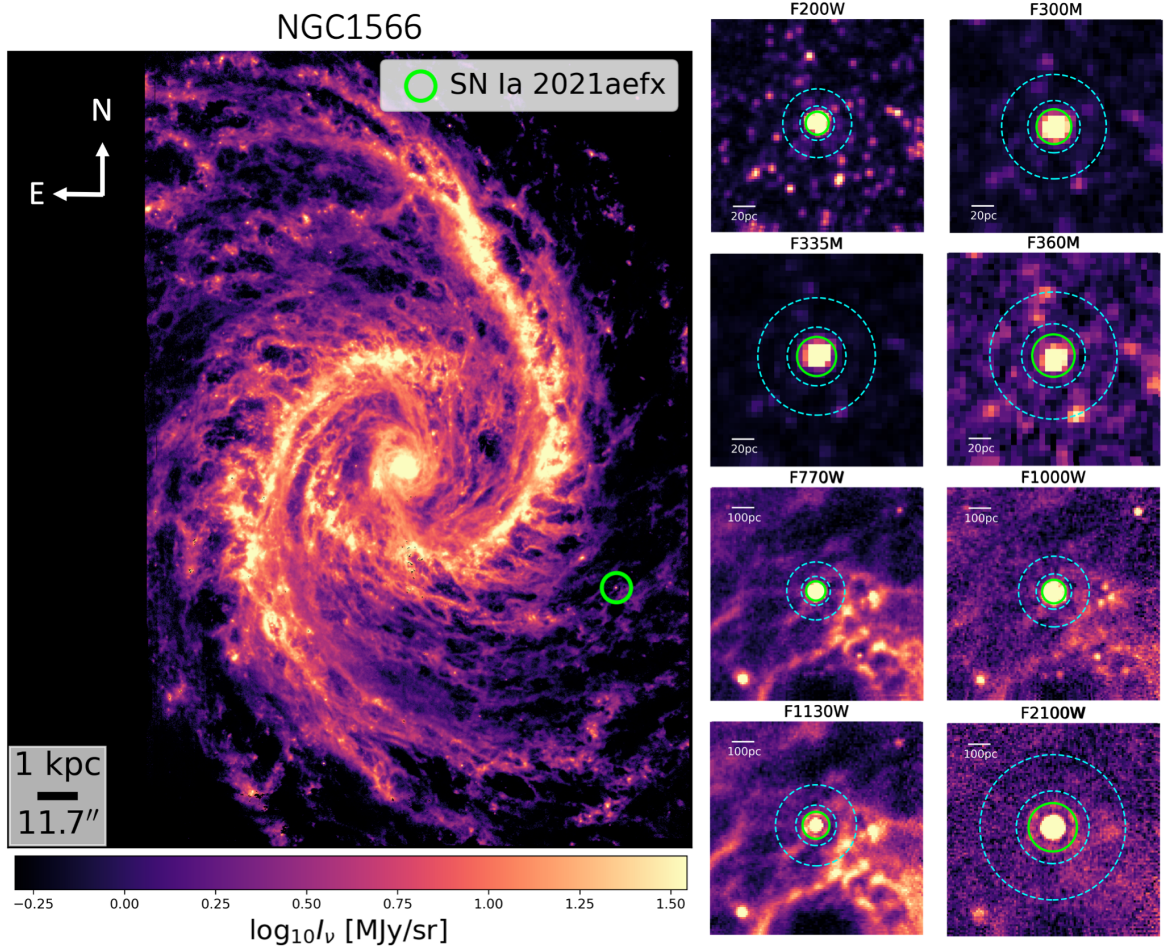}

\caption{\textit{\name\ in NGC~1566 at $\approx 2{-}21\mu$m.} Left panel: MIRI \emph{F1130W} PHANGS-JWST image of NGC~1566 showing the location of \name\ marked with a green circle. Right panels: zoom-ins on \name\ in each PHANGS-JWST filter. The top four panels show 200~pc~$\times~$200~pc cutouts from NIRCam images at 2.0$-$3.6 $\micron$. The bottom four panels show $1$~kpc~$\times$~$1$~kpc MIRI images at $7.7-21~ \micron$. The inner green circle marks the aperture used in the photometry measurement, and the two concentric, dashed, cyan circles mark the inner and outer apertures used for the background subtraction. The results of the photometry appear in Table \ref{tab:photometry}. \label{fig:SN2021aefx}}
\end{center}
\end{figure*}

\subsection{Near- and Mid-IR Photometry}

JWST \citep{Rigby2022} observed NGC~1566 on November 22, 2022 (MJD 59905.4) using both the near-infrared camera instrument \citep[NIRCam, ][]{Rieke2022} and mid-infrared instrument \citep[MIRI, ][]{Rieke2015, Wright2015} as part of the PHANGS-JWST Treasury survey \citep[GO program 2107, PI: J. Lee; ][]{Lee2022}. The observations include NIRCam imaging using the \emph{F200W}, \emph{F300M}, \emph{F335M}, and \emph{F360M} filters and MIRI imaging using the \emph{F770W}, \emph{F1000W}, \emph{F1130W}, and \emph{F2100W} filters. \citet{Lee2022} describe the general procedures for PHANGS-JWST observations and reduction. Here we use a preliminary version of the images reduced with the Calibration References Data System\footnote{\url {https://jwst-crds.stsci.edu/} }(CRDS) version 11.16.15 (NIRCam) and 11.16.16 (MIRI).


\begin{deluxetable*}{lcccccccclc}[]
\label{tab:photometry}
\tablecaption{Filter specific photometry data: exposure time, t; nominal wavelength at the filter center, $\lambda$; nominal wavelength width of the filter, $\Delta \lambda$; flux density, Flux; statistical uncertainty, $\sigma_{Stat}$; systematic uncertainty, $\sigma_{Sys}$; Luminosity times the frequency $\nu L_{\nu}$; and Luminosity times the nominal frequency width of the filter $\Delta \nu L_{\nu}$}
\tablehead{
\colhead{Filter} & 
\colhead{t} &
\colhead{$\lambda$} & \colhead{$\Delta \lambda$} & 
\colhead{Flux} & \colhead{$\sigma_{Stat}$} & 
\colhead{$\sigma_{Sys}$} & 
\colhead{$\nu L_{\nu}$}& 
\colhead{$\Delta \nu L_{\nu}$}\\
\colhead{}&
\colhead{[s]}&
\colhead{[$\micron$]}&
\colhead{[$\micron$]}&
\colhead{[$\mu$Jy]}&
\colhead{[$\mu$Jy]}&
\colhead{[$\mu$Jy]}&
\colhead{[$10^5 L_{\odot}$]}&
\colhead{[$10^5 L_{\odot}$]}&}
\startdata
F200W & 1203 & 1.99 & 0.461 & 20.12 & 1.55 & 0.85 & 3.012 & 0.707 \\
F300M & 387 & 2.996 & 0.318 & 14.2 & 0.15 & 1.45 & 1.412 & 0.15 \\
F335M & 387 & 3.365 & 0.347 & 19.28 & 0.21 & 1.67 & 1.707 & 0.176 \\
F360M & 430 & 3.621 & 0.372 & 4.77 & 0.2 & 0.2 & 0.392 & 0.04 \\
F770W & 88 &  7.7 & 2.2 & 268.52 & 1.39 & 26.34 & 10.388 & 3.03 \\
F1000W & 122 & 10.0 & 2.0 & 173.33 & 1.41 & 25.38 & 5.163 & 1.043 \\
F1130W & 311 & 11.3 & 0.7 & 181.41 & 2.3 & 11.24 & 4.782 & 0.297 \\
F2100W & 321 & 21.0 & 5.0 & 440.26 & 7.36 & 22.51 & 6.245 & 1.508 \\
\hline\hline
\enddata
\end{deluxetable*} 

Figure \ref{fig:SN2021aefx} shows the location of \name\ in NGC~1566 and presents cutouts around this location in each filter. The figure shows that in all eight filters, \name\ appears as a bright point source near the J2000 location of $\alpha = 64.9725^\circ$ and $\delta = -54.948081^\circ$ reported by \citet{Valenti2021}. Despite the significant variation of the PSF size from 2 to 21~$\mu$m, there is no issue identifying the location of the source in each band, and the SN always appears brighter than the surrounding emission.

We measure the flux of \name\ using aperture photometry and report our results in Table \ref{tab:photometry}. To do this, we use the aperture photometry module of the \textsc{astropy}-affiliated package \textsc{photutils}\footnote{\url{https://photutils.readthedocs.io/en/stable/aperture.html}}. Because we use initial versions of the PHANGS-JWST images in which the astrometry remains moderately uncertain, we center the circular aperture on the brightest pixel associated with the SN, which is slightly offset from filter to filter. We have no reason to think that the position of the SN varies by band and ascribe all of the slight variations in position to astrometric uncertainty. 

We assign aperture radii of 2 times the full width at half maximum (FWHM) of the point spread function (PSF) for each filter. We also estimate and subtract a local background using an annulus spanning from 1.5 to 3 times the radius of this photometric aperture (Figure \ref{fig:SN2021aefx}). Finally, we apply an aperture correction to each filter based on the fraction of total energy encircled within the PSF \citep[found using WebbPSF\footnote{\url {https://webbpsf.readthedocs.io/en/stable/index.html}};][]{Perrin2015} of our adopted aperture. 

We check our photometry by also estimating the flux using two independent methods. The first follows \citet[][]{Hoyer2022} and fits the source with two Sérsic profiles, one for the core and another for the halo of the source, without PSF convolution. The two profiles are integrated to find the total flux of the source, and a flat offset is added to account for the background. The second check uses the DOLPHOT stellar photometry package \citep{Dolphin2000, Dolphin2016}, which estimates the flux of sources using PSF-fitting applied to individual calibrated image frames (``level 2'') prior to drizzling the images into the final mosaics used for our primary analysis (``level 3''). Our DOLPHOT analysis uses the beta versions of the NIRCam\footnote{\url {http://americano.dolphinsim.com/dolphot/nircam.html}} \citep{Weisz2023} and MIRI\footnote{\url {http://americano.dolphinsim.com/dolphot/miri.html}}  \citep{Peltonen2023} modules. These measurements differ by an average of 6\% from our nominal
flux density measurement. 

We also estimate statistical and systematic uncertainties associated with the photometry. 

We use a relatively empty region of the image itself to assess the statistical uncertainty. To do this we focus on the northwest corner of the image, which is free of bright emission (Figure~\ref{fig:SN2021aefx}).
For each band, we generate 100 apertures with the same radius as the measurement aperture for that band, and place them randomly in the empty region. We then measure the standard deviation amongst the flux measurements for these empty apertures and treat this as a realistic estimate of the statistical uncertainty. Because both the local background and the source measurement varies across these ``empty'' apertures, we expect this estimate to reflect both uncertainty in the background level and noise within the aperture.

We include two contributions to the systematic uncertainty of each measurement, an overall flux calibration uncertainty and a methodological uncertainty. For the first, we assume a 4\% overall calibration uncertainty because JWST targets an overall flux calibration of 2\%\footnote{\url{https://jwst-docs.stsci.edu/jwst-data-calibration-considerations/jwst-data-absolute-flux-calibration}} and \cite{Rigby2022} quote a photometric reproducibility of 4\% for NIRCam. To be conservative, we assume that this 4\% value reflects the current calibration uncertainty of the telescope, and that this also applies to MIRI imaging. The second, methodological term reflects that different methods that ideally should yield identical results vary due to the influence of the background, imperfect knowledge of the PSF, or other methodological choices. To account for this, we take the median absolute deviation among the three flux measurements obtained with different methods for each filter, and then express this as a corresponding RMS uncertainty. Finally, we add both systematic uncertainties in quadrature and report them alongside the statistical uncertainty in Table \ref{tab:photometry}. 

Table \ref{tab:photometry} also reports $\nu L_\nu$ and $\Delta \nu L_\nu$, calculated by converting the flux density to a luminosity ($L_\nu$) and then multiplying by the frequency at the nominal filter center\footnote{\url {https://jwst-docs.stsci.edu/jwst-near-infrared-camera}} \footnote{\url {https://jwst-docs.stsci.edu/jwst-mid-infrared-instrument}}, $\nu$, and the nominal frequency width of the filter, $\Delta\nu$, respectively. $\Delta \nu L_\nu$ is a more useful physical measurement than $\nu L_\nu$ because it more nearly corresponds to a direct integral over the filter\footnote{This approach neglects second order effects related to the locations of the lines and the spectral slope adopted during calibration. When comparing to the \citet{Kwok2022} spectral data below we impute synthetic photometry from their spectra to allow a rigorous comparison.}, and the emission from the SN is expected to be localized line emission \citep[e.g., ][]{Fransson2015,Kwok2022}. 

\subsection{Flux Calibration of the +255~day Spectrum}

We compute synthetic photometry from the $0.3-14~\um$ +255~day spectrum of \name published by \citet{Kwok2022} so we briefly discuss the precision of the spectral flux calibration. \jwst is currently obtaining Cycle 1 calibration observations so the absolute flux calibration has not been finalized. However, there are several reasons why we consider the flux calibration of the +255~day spectrum to be sufficient. The optical portion of the +255~day spectrum has contemporaneous $UBgVri$ photometry to ensure precise flux calibration \citep{Kwok2022}. Additionally, the MIRI Low-Resolution Spectroscopy (LRS) spectrum is scaled by only $2\%$ to match the MIRI \emph{F1000W} acquisition-image photometry \citep{Kwok2022}. Thus, the optical and mid-IR wavelengths have flux calibration accurate to $\lesssim 5\%$. 

Only the NIRSpec observations lack corresponding photometry. While this could result in uncertain flux calibration for the $\approx 1-5~\um$ range, the NIRSpec spectrum shows excellent agreement with the flux-calibrated optical and mid-IR spectra on either end of the wavelength range. The current \jwst flux calibration is accurate to $2-5\%$ (S. Kendrew, private comm.) so we adopt a conservative uncertainty of $0.1$~mag for photometry synthesized from the NIRSpec ($1-5~\um$) portion and $0.05$~mag otherwise.

\subsection{Optical Photometry}

We supplement the \jwst observations with optical $BVgri$ imaging from the Precision Observations of Infant Supernova Explosions \citep[POISE; ][]{Burns2021} project. These observations were conducted on MJD~59904.7, $\approx 17$ rest-frame hours before the \jwst observations, with the SITe3 camera on the Swope telescope. Section 2 in \citet{Ashall2022} provides a brief overview of the data reduction and calibration process with further discussions in \citet{Krisciunas2017}. Finally, difference imaging with SkyMapper \citep{Wolf2018} data is used to mitigate the host-galaxy contribution although the effect is minimal for \name as it is located in the outskirts of NGC~1566.


\section{Results}
\label{sec:Results}


\begin{figure*}
    \centering 
    \includegraphics[width=0.99\linewidth]{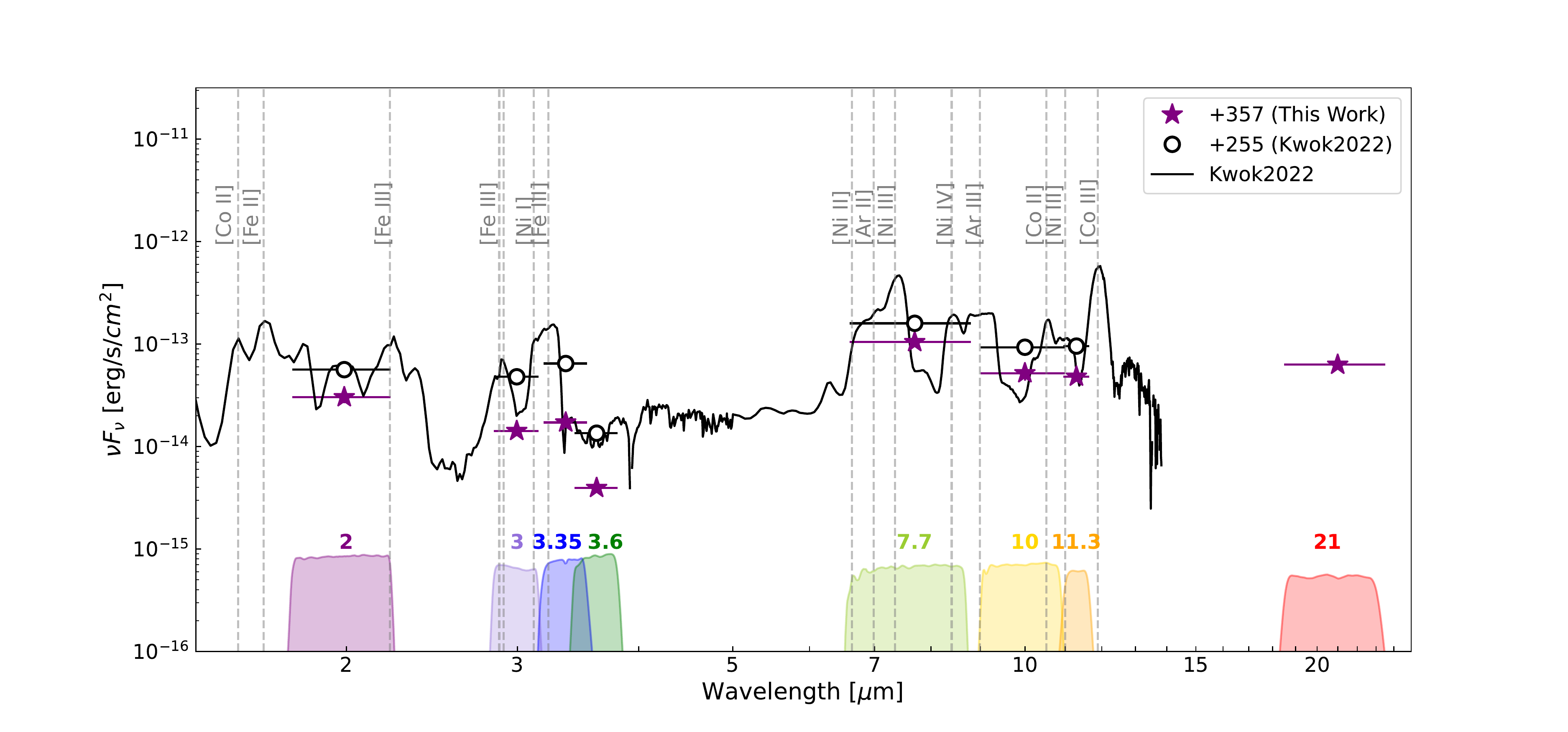}
\caption{\textit{A comparison of the Near- and Mid- IR flux of \name\ at +255 and +357 days.} Purple stars mark our flux measurements from each filter and  black open circles signify synthetic photometry from the +255-day spectrum (black line). Horizontal error bars show the nominal wavelength width of each filter. Vertical error bars mark the uncertainty in the measurement and are shown, but are typically smaller than the points. The vertical grey dashed lines mark emission features identified by \cite{Kwok2022}. NIRCam and MIRI filter curves are plotted along the bottom axis.
\label{fig:SED}}
\end{figure*}

Figure~\ref{fig:SED} shows our new near- and mid-IR photometry of \name at $+357$~days after maximum light. Our new measurements appear as purple stars, with the filters illustrated below the SED and the nominal filter widths indicated by the horizontal error bars. Filter curves are taken from the Spanish Virtual Observatory's filter profile service\footnote{\url{http://svo2.cab.inta-csic.es/theory/fps3/}} \citep{Rodrigo2020}. The solid line behind our data shows the spectrum of \citet{Kwok2022}. For comparison, we also compute synthetic photometry from the +255~day $0.3-14~\um$ spectrum of \name \citep{Kwok2022}, which appears as black-and-white circles.

Comparing the synthesized photometry to the new \jwst observations, one can see that the shape of the SED measured from our data resembles that found by \citet{Kwok2022}. While \name has faded slightly during the $\approx$ 102~days between these observations, the dominant emission features appear to have similar strengths with the exception of Co-dominated features \citep{derkacy2023} as $^{56}$Co decays into $^{56}$Fe. This is consistent with the optical and near-IR spectroscopic evolution of \sneia during this time span \citep[e.g., ][]{Mazzali2020, Tucker2022a, Graham2022}. 

Our \emph{F2100W} detection is unique because it was not covered by the spectrum obtained by \citet{Kwok2022}. No models in the literature currently predict which emission features should be present at these wavelengths. \citet{Fransson2015} suggest that several MIR fine-structure lines dominate the total emission of SN~2011fe at $\sim 1000$~days after \tmax, but time-dependent modeling is unavailable. Thus, we use simple assumptions to estimate the emission features responsible for the observed \emph{F2100W} flux. 

\begin{table*}
    \centering
    \caption{Integrated fluxes $F$ and fractional contribution to the total $0.3-14~\um$ flux $f$ for different wavelengths. Fractions may not exactly equal 100\% due to round-off and sampling errors.}
    \begin{tabular}{ccrcrc}
    \hline\hline
        $\lambda$ & $F_{255}$ & $f_{255}$ & $F_{357}$ & $f_{357}$ & $\Delta m$\\
           \ [$\um$]    & [erg~s$^{-1}$~cm$^{-2}$]    & [\%] &  [erg~s$^{-1}$~cm$^{-2}$] & [\%] & [mag/100~days] \\\hline
        $0.3-1$ & $(1.3\pm0.1)\times10^{-12}$ & $86.1\pm3.9$ & $(3.0\pm0.2)\times10^{-13}$ & $72.0\pm4.6$ & $1.57\pm0.04$\\
        $1-2.5$ & $(6.9\pm0.3)\times10^{-14}$ & $4.7\pm0.3$ & $(4.6\pm0.4)\times10^{-14}$ & $11.3\pm1.0$ & $0.43\pm0.08$ \\
        $2.5-5$ & $(2.1\pm0.1)\times10^{-14}$ & $1.4\pm0.1$ & $(5.6\pm0.8)\times10^{-15}$ & $1.4\pm0.2$ & $1.42\pm0.15$ \\
        $5-14$ & $(1.1\pm0.1)\times10^{-13}$ & $7.5\pm0.4$ & $(6.0\pm0.5)\times10^{-14}$ & $14.9 \pm1.4$ & $0.63\pm0.08$ \\
        $0.3-14$ & $(1.5\pm0.1)\times10^{-12}$ & $\ldots$ & $(4.1\pm0.2)\times10^{-13}$ & $\ldots$ & $1.35\pm0.05$ \\
        \hline\hline
    \end{tabular}
    \label{tab:fluxes}
\end{table*}

\begin{figure*}
    \centering
    \includegraphics[width=\linewidth]{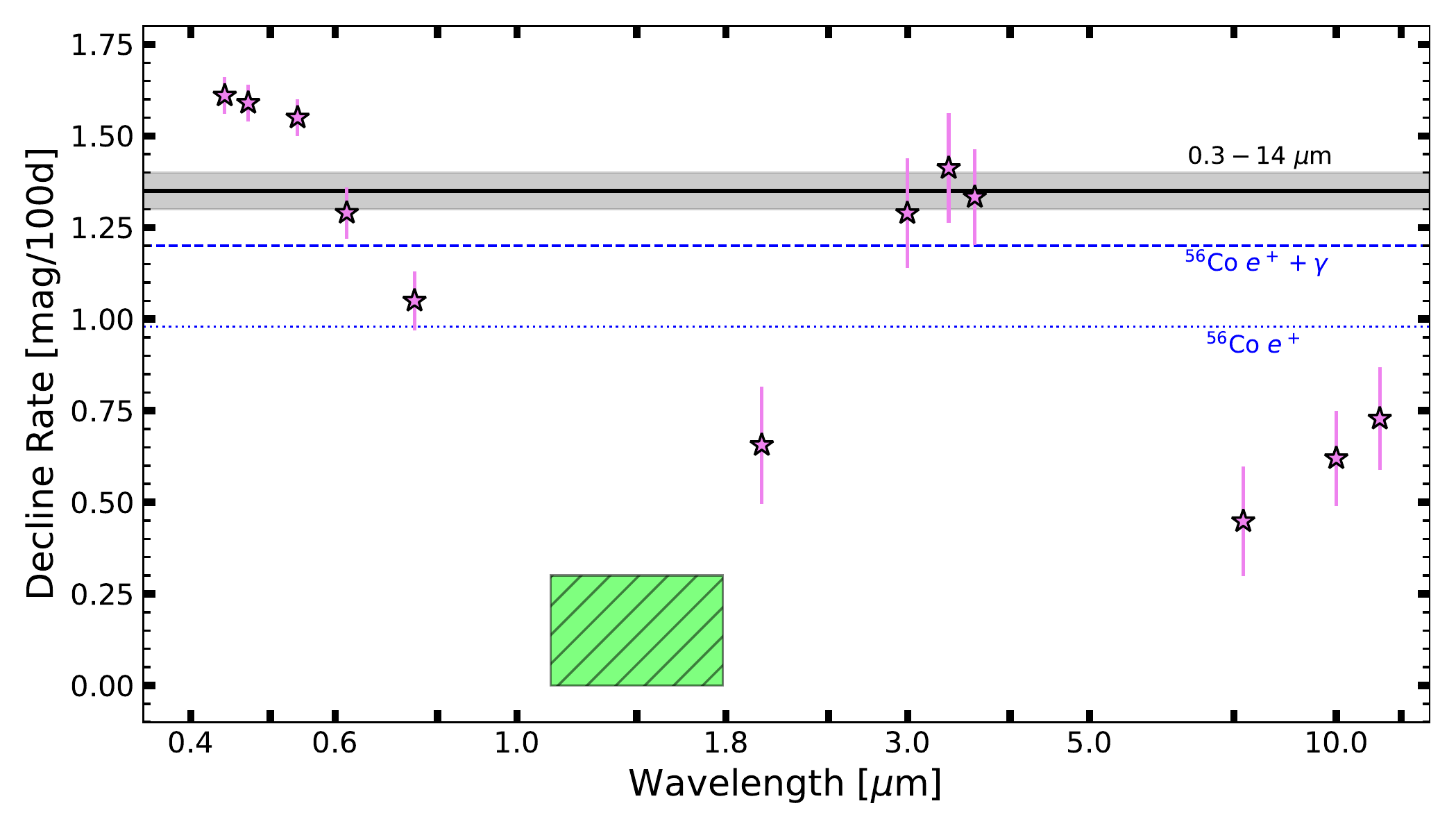}
    \caption{Nebular-phase decline rates as a function of wavelength for \name. The green region represents the $J$- and $H$-band decay rates of $0.0-0.3$~mag/100~days from \citet{Graur2020}. Blue horizontal lines show the energy deposition rate for $^{56}$Co with (dashed) and without (dotted) including the $\gamma$-ray contribution. The black line and shaded region represent the integrated $0.3-14~\um$ decline rate derived in \S\ref{sec:Results}. Data for this figure are included in the online version of the manuscript.}
    \label{fig:nebdec}
\end{figure*}

\begin{figure}
    \centering
    \includegraphics[width=\linewidth]{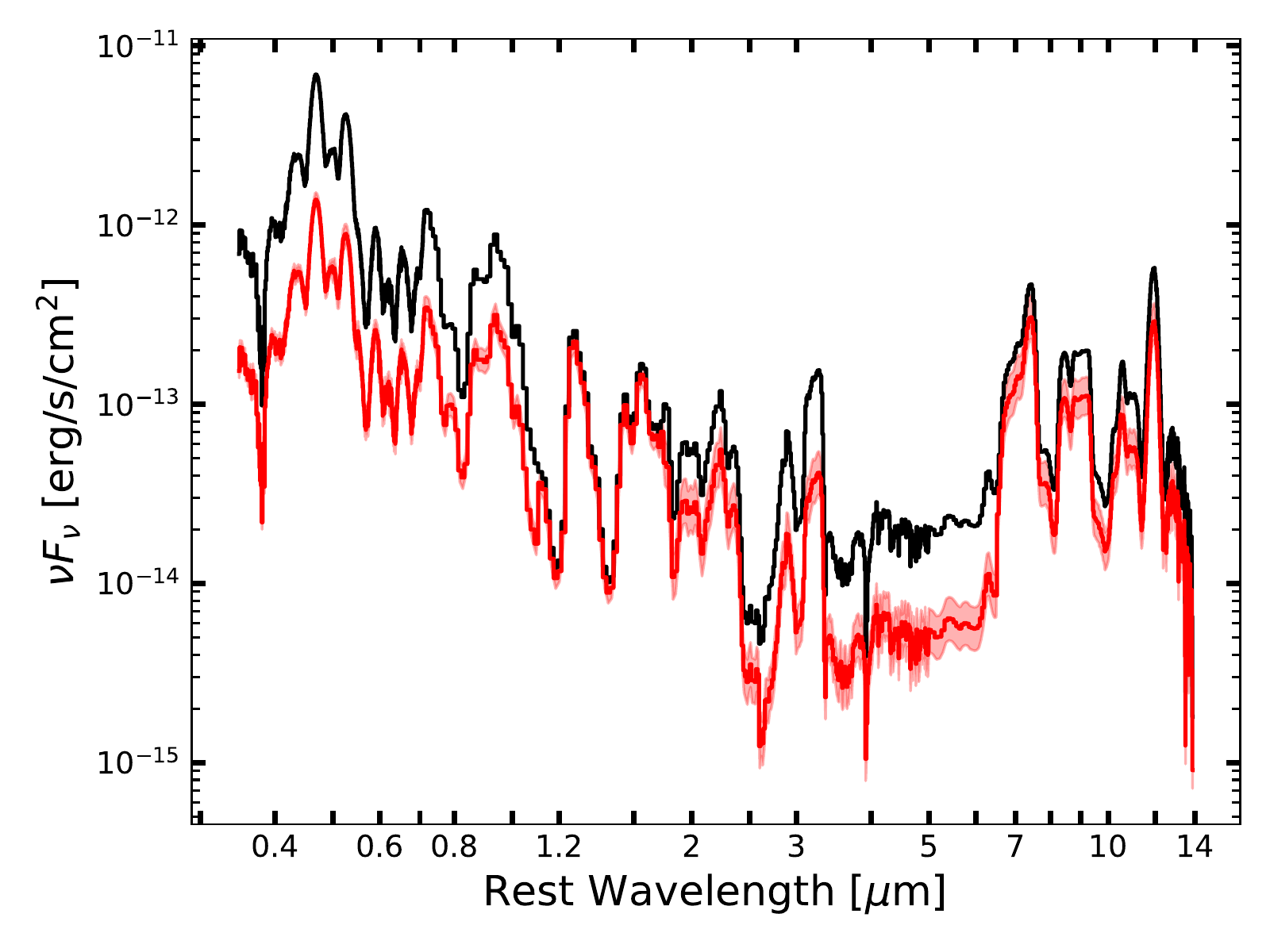}
    \caption{Flux-calibrated +255~day spectrum from \citet{Kwok2022} (black) and our reconstructed +357~day spectrum (red). The shaded region represents the 95\% confidence interval from the bootstrapping routine.}
    \label{fig:warpspec}
\end{figure}

The temperature and electron density at these epochs are $T_e\approx10^4$~K and ${\log_{10}(n_e [\rm{cm}^{-3}])\approx 5-6}$ (e.g., \citealp{Fransson2015, Flors2020, Shingles2022}). The ground-state [\ion{Fe}{3}]~$22.93~\um$ transition likely accounts for most of the observed flux despite the \emph{F2100W} bandpass capturing only $\sim 50\%$ of the total emission. The rest of the observed flux might be attributable to a combination of the weaker [\ion{Ar}{3}]~$21.83~\um$ and [\ion{Co}{2}]~$18.80~\um$ lines, although detailed modeling is needed to accurately capture the complex non-LTE processes at nebular epochs \citep[e.g., ][]{Li2012, Shingles2020, Hoeflich2021}. Future observations with the \jwst MIRI Mid-Resolution Spectrometer \citep[MRS; ][]{Wells2015} will directly measure the contributing spectral features. 

Combining the flux-calibrated +255~day spectrum with our new optical and IR photometry, we estimate the fraction of flux emerging in the optical, near-IR, and mid-IR for both epochs. The \emph{F2100W} measurement is excluded from these calculations due to the uncertain spectral shape as \sneia are dominated by line emission at nebular epochs. The +255~day spectrum is scaled to match the observed +357~day optical and IR photometry using the filter-specific decline rates shown in Figure!~\ref{fig:nebdec}. A bootstrapping routine is used to estimate the associated uncertainty. For the $1-2~\um$ region where we lack corresponding photometry, the $J$- and $H$-band decline rate is uniformly sampled between $0.0-0.3$~mag/100~days which was found for a sample of \sneia by \citet{Graur2020}. The final scaled spectrum and its uncertainty are shown in Figure~\ref{fig:warpspec}. The fraction of $1-2.5~\um$ to $0.3-1~\um$ flux is $5.1\pm0.5\%$ and $13.4\pm1.3\%$ at +255 and +357~days, respectively, in agreement with previous estimates for other \sneia \citep{Stanishev2007, Leloudas2009, Dimitriadis2017, Maguire2018, Graur2020}. We consider this agreement, without adopting prior assumptions on the optical to near-IR flux ratios, a general confirmation that our methodology is accurate within our quoted uncertainties.

Table~\ref{tab:fluxes} provides the fluxes for each wavelength region in addition to the fraction of the total $0.3-14~\um$ flux. The majority of the total flux is emitted at $<1~\um$ for both epochs, although the relative fraction decreases from $f_{\rm{opt}}\approx 85\%\rightarrow70\%$. This decrease is compensated by the increasing fractional flux at $1-2.5~\um$ and $5-14~\um$. The $2.5-5~\um$ region appears to retain the same fraction of the total flux between the two epochs, although this spectral region only consists of $\approx 1.5\%$ of the total flux. 

Finally, we use the integrated fluxes to compute the decline rates for the different wavelength regions listed in Table~\ref{tab:fluxes}. We find that the majority of the emitted energy occurs at optical wavelengths, and this wavelength region also declines the fastest due to higher-energy transitions becoming harder to populate as the density decreases. The $\approx 3-4~\um$ region also declines quickly due to the presence of [\ion{Fe}{3}] emission features. Spectral regions corresponding to singly-ionized transitions ($1-2.5~\um$, $6-12~\um$) decline slower at $\lesssim 0.6$~mag/100~days, in agreement with the findings of \citet{Gerardy2007} and \citet{McClelland2013}. However, the fast-declining optical wavelengths dominate the total flux producing a total $0.3-14~\um$ decline rate of $\Delta m_{0.3-15} = 1.35\pm0.05$~mag/100~days. This is higher than the decay rate of 0.98~mag/100~days from $^{56}$Co positrons. Including the small amount of heating from singly-scattered $\gamma$-ray photons \citep[e.g., Eq. 1 from ][]{Kushnir2020} raises the predicted decline rate to 1.20~mag/100~days which still disagrees with our observed decline rate at $\sim 3\sigma$.

\section{Discussion}
\label{sec:Discussion}

We have provided new nebular-phase optical and IR imaging of the SN Ia 2021aefx, including the first detection of a \snia at $>15~\um$. Only a few \sneia have mid-IR observations so comparisons are limited. However, the general picture outlined by \citet{Fransson2015} and \citet{Graur2020} appears robust. In this picture, ultraviolet flux from Fe transitions is redistributed into the optical and IR via multiple scatterings and fluorescence \citep[e.g., ][]{Pinto2000}. Spectral regions corresponding to doubly-ionized transitions typically decay faster than their singly-ionized counterparts due to the ejecta cooling as it expands \citep[e.g., ][]{McClelland2013}. However, this framework for interpreting the observed decline rates may be overly simplistic as the pseudo-continuum is produced by a multitude of overlapping spectral features \citep[e.g., ][]{Fransson2015, Mazzali2015, Wilk2020}, which may or may not evolve similarly to the strongest emission feature in a given bandpass. 

The derived (pseudo-)bolometric decay rate of $1.35\pm0.05$~mag/100~days is higher than expected for $^{56}$Co decay. The most likely explanation for this discrepancy is that more flux is emerging at $>14~\um$ at later epochs. [\ion{Fe}{2}] has two low-energy transitions at $24.52~\um$ and $25.99~\um$ with lower-level energies of $E_l = 0.30$~eV and $E_l = 0.00$~eV, respectively. The nebular-phase $J$- and $H$-band plateaus are driven by low-energy [\ion{Fe}{2}] transitions at $1-2~\um$, so it is plausible that the mid-IR transitions experience a similar flattening in the decline rate. If there is no non-radiative energy loss at these epochs (i.e., the bolometric light curve follows the energy generation rate), then the flux at $>14~\um$ must have a decline rate of $\approx 0.1$~mag/100~days to reconcile our results with predictions for $^{56}$Co decay. 

Although we find flux emerging at $>14~\um$ the most plausible explanation, there are other physical processes that can expedite the (pseudo-)bolometric decay relative to $^{56}$Co. Allowing a few percent of the high-energy positrons to escape the ejecta into the surrounding environment will increase the nebular-phase decline rate. In this scenario, the bolometric light curve does not reflect the energy deposition rate as some of the radioactive decay energy is not converted into photons. Magnetic fields can restrict the propagation of positrons within the ejecta but the source and strength of such fields remain unclear \citep[e.g., ][]{Hristov2018}.
Furthermore, the fraction of escaping positrons would need to increase between +255 and +357~days, which is possible if the magnetic field strength diminishes as the density decreases. Allowing even a small fraction of positrons to escape the ejecta would have important ramifications for our understanding of the 511~keV positronium annihilation signal observed in the Milky Way \citep[e.g., ][]{Siegert2016}. Conversely, complete positron confinement requires a relatively strong ($\gtrsim 10^6$~G) and/or morphologically complex magnetic field \citep[e.g., Table 2 from ][]{Hristov2021}. Strong magnetic fields can be produced in the merger of two white dwarfs \citep[e.g., ][]{Williams2022} but it is unclear how the magnetic field strength and morphology will be affected by the explosion \citep[e.g., ][]{Remming2014}. Alternatively, particle streaming instabilities can generate magnetic fields \citep[e.g., ][]{Gupta2021}, but no simulations have assessed if nebular-phase \sneia ejecta meet the required instability criteria.  

Separately, \citet{Kushnir2020} propose that the expansion of the ejecta will extend the time over which $^{56}$Co positrons deposit their energy into the ejecta. This ``delayed deposition'' effect should result in a steepening of the light curve (relative to $^{56}$Co decay) at $\lesssim500$~days after \tmax due to the finite time needed for positrons to deposit their energy into the ejecta. The deviation from $^{56}$Co decay is only a few percent of the energy generation rate at $\sim 200-400$~days after \tmax which can qualitatively reproduce our pseudo-bolometric decline rate.  However, these epochs correspond to a flattening in the differential heating rate (see Figure~4 in \citealp{Kushnir2020}) which may negate any deviations from $^{56}$Co decay during our observational baseline.

Future observations of \sneia with \jwst are crucial to better understand the complex physics governing \sneia at nebular phases. Multi-epoch photometry constrains the mass ratios of isotopes produced during high-density burning \citep[e.g., ][]{Shappee2017, Kerzendorf2017}. The late-time $(\gtrsim 500$~days after \tmax) mid-IR luminosity is a key diagnostic of the ejecta's cooling properties, as these epochs correspond to the end of the $H$-band light curve plateau \citep{Graur2020} and a distinct shift in the optical ionization properties \citep{Tucker2022a}. Several lines of evidence suggest clumping is present in \snia ejecta \citep[e.g., ][]{Black2016, Mazzali2020} but specifics about the physical conditions remain elusive \citep[e.g., ][]{Wilk2020, Shingles2022}. Emission-line profiles in the near- and mid-IR should be able to directly ascertain the nebular-phase magnetic field strength \citep{Penney2014} and further constrain positron propagation (or lack thereof) within the ejecta \citep[e.g., ][]{Hristov2021}. 

\jwst is poised to revolutionize our understanding of \sneia in the nebular phase with both dedicated \snia programs (e.g., GO 2072, PI: Jha; GO 2114, PI: Ashall) and serendipitous observations like those presented here. To date, the PHANGS-JWST program has observed six nearby star-forming galaxies (IC~5332, NGC~0628, NGC~1365, NGC~1385, NGC~1566, NGC~7496). Of those galaxies, two have confirmed \sneia in the \jwst footprint that exploded within the past 20~years: SN~2010el \citep[SN Iax,][]{Foley2013} and SN~2021aefx in NGC~1566 and SN~2012fr in NGC~1365. SN~2021aefx is analyzed here. SN~2010el is located coincident to heavy background emission in the MIRI filters and is not visible in any of the available filters (e.g., Table~\ref{tab:photometry}). Similarly, there is no obvious emission seen at the location of SN~2012fr. These results are unsurprising because the \jwst images were obtained 9.75 years after \tmax \citep{Contreras2012} for SN~2012fr and $\sim$12.5 years after the discovery \citep{Monard2010} of SN~2010el. 

The PHANGS-JWST Cycle 1 Treasury Program will be observing 13 more galaxies in addition to those listed above and we will continue searching for SN emission in future PHANGS data. Even the SNe that do not show clear nebular-phase emission offer an unprecedented, sharply resolved view of the environments into which SNe deposit their energy and momentum. Looking at the IR emission at the locations of these SNe can be used to constrain the late-time impact of these explosions and assess their role in the evolution of the ISM \citep[e.g., ][]{MaykerChen2022}.


\vspace{1cm}
\textit{Facilities}: \emph{JWST} (NIRCam, MIRI), Swope (SITe3)

\textit{Software}: Astropy \citep{Astropy2013}, photutils \citep{PhotUtils2020}, Jupyter \citep{Kluyver2016}


\section*{Acknowledgments}

This work was carried out as part of the PHANGS collaboration.

AKL and NMC gratefully acknowledge support by grants 1653300 and 2205628 from the National Science Foundation, award JWST-GO-02107.009-A, award SOSP SOSPADA-010 from the NRAO, and by a Humboldt Research Award from the Alexander von Humboldt Foundation.

ER acknowledges the support of the Natural Sciences and Engineering Research Council of Canada (NSERC), funding reference number RGPIN-2022-03499 and the support of the Canadian Space Agency (CSA) [22JWGO1-
20].

JD, CA, and PH acknowledge support by NASA grant
JWST-GO-02114.032-A.

MB acknowledges support from FONDECYT regular grant 1211000 and by the ANID BASAL project FB210003.

FB acknowledges funding from the European Research Council (ERC) under the European Union’s Horizon 2020 research and innovation programme (grant agreement No.726384/Empire).

OE gratefully acknowledge funding from the Deutsche Forschungsgemeinschaft (DFG, German Research Foundation) in the form of an Emmy Noether Research Group (grant number KR4598/2-1, PI Kreckel). 

ES acknowledges funding from the European Research Council (ERC) under the European Union’s Horizon 2020 research and innovation programme (grant agreement No. 694343).

L.G. acknowledges financial support from the Spanish Ministerio de Ciencia e Innovaci\'on (MCIN), the Agencia Estatal de Investigaci\'on (AEI) 10.13039/501100011033, and the European Social Fund (ESF) "Investing in your future" under the 2019 Ram\'on y Cajal program RYC2019-027683-I and the PID2020-115253GA-I00 HOSTFLOWS project, from Centro Superior de Investigaciones Cient\'ificas (CSIC) under PIE 20215AT016 and LINKA20409 projects, and the program Unidad de Excelencia Mar\'ia de Maeztu CEX2020-001058-M.

M.O. acknowledges support from UNRN PI2020 40B885.

M.D. Stritzinger is funded by the Independent Research Fund Denmark (IRFD, grant number 10.46540/2032-00022B ).

This work is based on observations made with the NASA/ESA/CSA James Webb Space Telescope. The data were obtained from the Mikulski Archive for Space Telescopes at the Space Telescope Science Institute, which is operated by the Association of Universities for Research in Astronomy, Inc., under NASA contract NAS 5-03127 for JWST. These observations are associated with programs \#02107 and \#02072. The specific observations for PHANGS-JWST can be accessed via \dataset[ 10.17909/9bdf-jn24]{http://dx.doi.org/10.17909/9bdf-jn24}.

\vspace{3cm}

\bibliography{main.bbl}



\end{document}